\documentclass[aps,prl,twocolumn,groupedaddress,showpacs]{revtex4}
\usepackage{graphicx}

\begin{document}

\title{The complex scaling behavior of non--conserved self--organized
critical systems}


\author{Barbara Drossel}
\affiliation{Institut f\"ur Festk\"orperphysik,  TU Darmstadt,
Hochschulstra\ss e 6, 64289 Darmstadt, Germany }

\date{\today}

\begin{abstract}
The Olami--Feder--Christensen earthquake model is often considered  the
prototype dissipative self--organized critical model. It is shown that
the size distribution of events in this model results from a complex
interplay of several different phenomena, including limited
floating--point precision. Parallels beween the dynamics of synchronized
regions and those of a system with periodic boundary conditions are pointed
out, and the asymptotic avaanche size distribution is conjectured to
be dominated by avalanches of size one, with the weight of larger
avalanches converging towards zero as the  system size increases.
\end{abstract}


\maketitle

Self--organized critical (SOC) systems \cite{bak87,bakbuch} are
extended systems that receive a slow energy input and have fast
dissipation events (``avalanches'' or ``earthquakes''), the size
distribution of which is a power law. Such a scale invariance can
result quite naturally in systems that satisfy a local conservation
law \cite{hwa89}, however, the mechanisms leading to SOC in
non--conserved systems are not yet well understood.  The
Olami--Feder--Christensen earthquake model \cite{ofc92} is probably
the most studied nonconservative SOC model. Nevertheless, the nature
of its critical behavior is still not clear.  Despite the simplicity
of its dynamical rules, the model shows a variety of interesting
features that are unknown in equilibrium physics and appear to be
crucial for generating the critical behavior. Among these features are
a marginal synchronization of neighboring sites driven by the open
boundary conditions \cite{mid95}, and the violation of finite--size
scaling \cite{gras94,lis01} together with a qualitative difference
between system--wide earthquakes and smaller earthquakes
\cite{lis01a}. Also, small changes in the model rules (like replacing
open boundary conditions with periodic boundary conditions
\cite{per96}, or introducing frozen noise \cite{mou96}), destroy the
SOC behavior.

The model is a discretized and simplified version of the
Burridge--Knopoff model of earthquakes \cite{bur67} and is defined by
the following rules: At each site of a square lattice, a continuous
variable $z$ is defined that represents a local force. The force at
all sites increases uniformly at constant rate. When the force
$z(i,j)$ at a site $(i,j)$ exceeds the threshold value $z_c$, which
can be chosen to be $z_c=1$ without loss of generality, the force at
this site is reset to zero, while all four nearest neighbors receive a
force increment of $\alpha z(i,j)$. The parameter $\alpha$ is the only
parameter of the model, and it has a value in the interval
$(0,0.25)$. If a neighbor is lifted above the threshold, the force on
its neighbors is increased according to the same rule, etc., until the
``earthquake'' is finished. The ``size'' of an earthquake, $s$, is
defined to be the number of toppling events during this
earthquake. Then, the force is again increased uniformly, until the
next site reaches the threshold, triggering the next earthquake, and
so on.

Similarly to real earthquakes, which follow the Gutenberg--Richter
law, the size distribution of earthquakes (which we shall subsequently
call ``avalanches'') in this model is found in computer simulations to
resemble a power law $n(s) \sim s^{-\tau}$. The exponent $\tau$
appears to depend on $\alpha$. However, recent evidence points to a
universal exponent $\tau \simeq 1.8$ if only the system size is made
large enough, at least for $\alpha \ge 0.17$ \cite{lis01}.  Several
authors argue that there is a critical value of $\alpha$, around 0.18,
below which the dynamics change qualitatively and become dominated by
small avalanches \cite{gras94,bot97}.  There is possibly a small value
of $\alpha$, around 0.05, below which the power law breaks down
completely \cite{ofc92} (however, this idea has few supporters now),
and one recent publication even claims that the model is not critical
at all for $\alpha < 0.25$ \cite{car00}. This latter claim is based on
the finding that the mean branching ratio of avalanches is smaller
than one, and it would be correct if there was only one type of
avalanches in the system.

It is the purpose of this paper to clarify some of these puzzles and
to elucidate the mechanisms that generate the observed avalanche size
distributions. We will focus on the cases $\alpha=0.1$ and
$\alpha=0.05$ and show that the avalanche size distribution results
from the complex interplay of several phenomena, including boundary
driven synchronization and internal desynchronization, limited
floating-point precision, the slow dynamics within the steady state,
and the small size of synchronized regions. While part of these
phenomena have been pointed out already in the earlier literature,
their combined effect on the size distribution of earthquakes and
their implications for the asymptotic scaling behavior have not been
clarified so far. It is our prediction that the asymptotic avalanche
size distribution (in the ideal situation of infinite floating-point
precision) is dominated by avalanches of size 1, with the weight of
larger avalanches decreasing to zero with increasing system size.

\begin{figure}
\includegraphics[width=5cm]{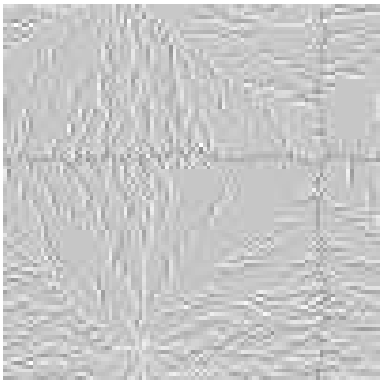}
\caption{Snapshot of the periodic state for $L=100$ and $\alpha=0.1$,
with an initial width 0.2 of the distribution of $z$ values. The force
$z$ is visualized by the grey shade, with smaller force being darker.
\label{snapper}}
\end{figure}

\begin{figure}
\includegraphics[width=5cm]{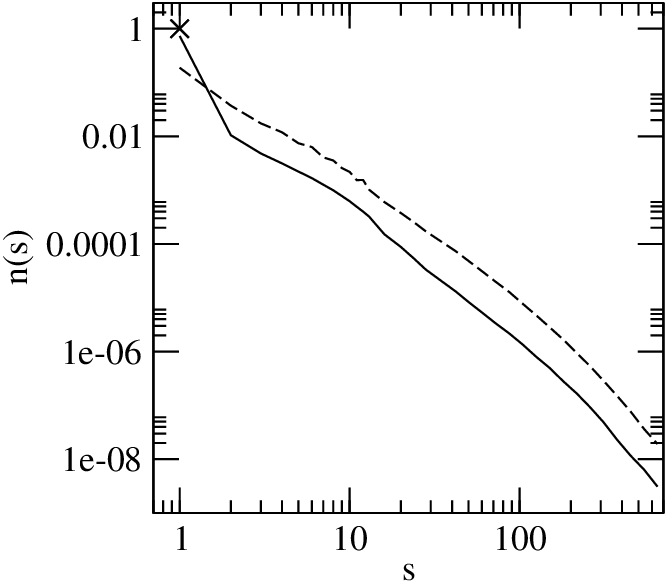}
\caption{Size distribution of avalanches in the periodic state for
$L=100$ and $\alpha=0.1$, with an initial width 0.2 of the
distribution of $z$ values. Dashed line: ``Float'' precision. Solid line: ``Double'' precision. Cross: Projected result for infinite precision.
\label{avalper}}
\end{figure}

Let us first discuss the case of periodic boundary conditions, which
will be relevant also for the later discussion of the system with open
boundary conditions. Starting with a random initial configuration
(most authors choose the $z$ values randomly from the interval
$[0,1)$), the dynamics settle on a periodic attractor, where the sites
always topple in the same order \cite{mid95,gras94,soc93}. Often the
sites are found to topple one by one, without any avalanches of size 2
or larger. Periodic states with larger avalanches were found for
values $\alpha>0.18$ \cite{gras94}, and they occur also for smaller
$\alpha$ when the width of the initial distribution of $z$ values is
made smaller. However, Middleton and Tang have shown analytically
\cite{mid95} that a system consisting of two sites always settles in a
periodic state where none of the two sites induces the toppling of the
other, leading to the expectation that this should also hold for two
sites that are part of a system with periodic boundary conditions in a
periodic stationary state. In fact, a closer inspection of the
periodic states with larger avalanches reveals that within an
avalanche sites are always lifted exactly to the threshold value by
the force increment they receive from the toppling neighbor.  At this
point, the process pointed out by Middleton and Tang comes to a halt
due to the finite floating--point precision, because each site in the
system receives during each period the same force increment of
$4\alpha$ from its neighbors, allowing thus for a periodic state. If
the precision was infinitely large, the periodic state could only be
reached if no site induced the toppling of another site.  I tested
this conclusion by studying a system of linear size $L=100$ with
$\alpha=0.1$ and with the initial values of $z$ randomly chosen from
an interval of size $2\alpha$. The first avalanche covers the entire
system, but subsequent avalanches become smaller, until a periodic
state is reached.  Figure \ref{snapper} shows a snapshot of the
periodic state, just before the first site topples.
The pattern visible in this snapshot was generated during the first,
system-wide avalanche. The site that topples first during a cycle has
the largest force. The bright (but not completely white) sites topple
after one of their nearest neighbors has toppled. The light grey sites
(the majority) topple after two of their neighbors have toppled, and
the remaining ones topple only after three or four neighbors have
toppled. The force difference between two neighbors has therefore
peaks at multiples of $\alpha$, as already found by Grassberger for
larger values of $\alpha$ \cite{gras94}.  Figure \ref{avalper} shows
the avalanche size distribution in the periodic state, averaged over
many different realisations of the disorder.  $n(s)$ is the total
number of avalanches of size $s$, divided by the number of sites and
by the number of cycles. A cycle corresponds to increasing the force
by $1-4\alpha$. In a periodic state each site returns
to its original force value at the end of a cycle. 
For larger precision (``double'' instead of ``float''), there are more
smaller and less larger avalanches, supporting the claim that the
floating--point precision limits the degree of
desynchronization. Within an avalanche, all sites are lifted exactly
to the threshold (within the given precision), justifying the
prediction that for infinite precision all avalanches are of size 1. 

Middleton and Tang have also shown that if one of two sites has a
somewhat shorter period than the other one (for instance by having a
lower threshold value or a smaller $\alpha$), then the two sites
settle in a periodic state where the site with the longer period
always induces toppling of the site with the shorter
period. Similarly, systems with periodic boundary conditions and
frozen disorder in the threshold values or in the values of $\alpha$
can have a synchronized state where all sites topple during the same
avalanche, if the strength of the disorder is not too large. (This is
shown in \cite{mou96} for disorder in the $\alpha$ values. The
strength of disorder in the threshold values in \cite{jan93} was too
large to see the synchronization, but I have observed synchronization
for weaker disorder.)

\begin{figure}
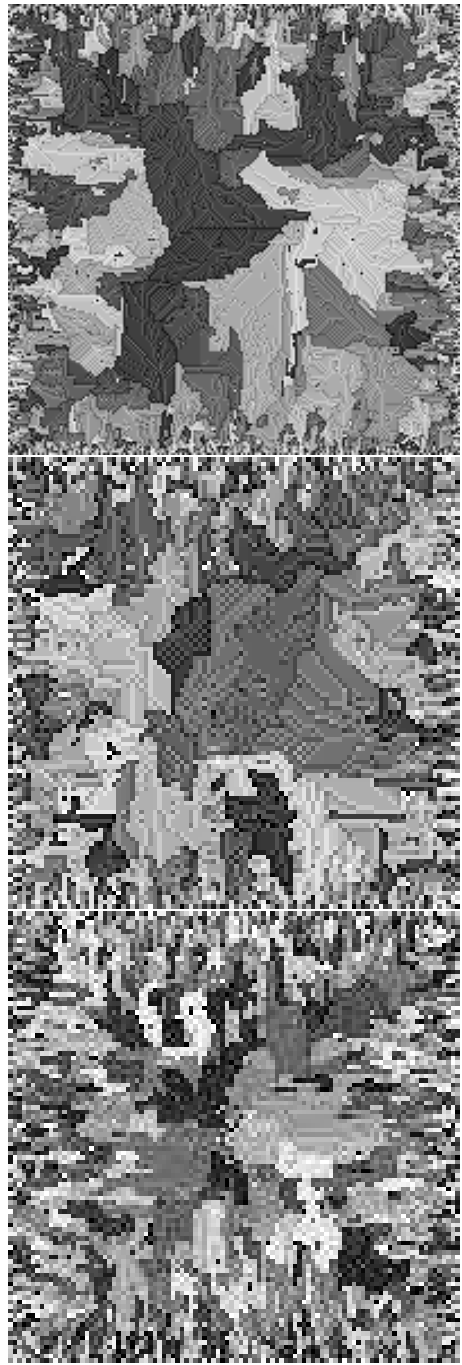

\includegraphics[width=6cm]{fig3a.eps}
\includegraphics[width=6cm]{fig3b.eps}
\includegraphics[width=6cm]{fig3c.eps}
\caption{Snapshots of the stationary state for $L=200$ and
$\alpha=0.1$ (top), $L=100$ and $\alpha=0.1$ (center) and $L=100$ and
$\alpha=0.05$ (bottom). The force $z$ is visualized by the grey
shade, with smaller force being darker. 
\label{snaps}}
\end{figure}

When the system has open boundary conditions, sites at the boundary
receive less force from their neighbors, and can therefore be
considered as having a smaller period, according to Middleton and Tang
\cite{mid95}.  This explains the observed formation of synchronized
regions, which starts from the boundary and proceeds inward with time,
apparently according to a power-law \cite{mid95}, which seems to be
universal for $\alpha \ge 0.15$ \cite{lise02}.  After some time (which
may be extremely long for large $L$ or small $\alpha$), the system
reaches a stationary state with a ``patchy'' structure. We will call a
patch a ``synchronized region''. Within a synchronized region, the
sites have similar $z$ values and topple within a short time of each
other, usually through a sequence of small avalanches. Regions that
are further away from the boundary are larger. The regions change
their shape and size on a time scale that is much longer than the
timescale of driving, due to large avalanches that are triggered close
to the boundary.  For smaller $\alpha$, there are more and smaller
such regions, while for $\alpha$ larger than around 0.17, most of the
system is dominated by one or a few large regions. This paper focusses
on the case of smaller values of $\alpha$, where there are several
regions of different sizes, and where the uncertainty about the
model's behavior is largest. The results reported in the following
were obtained for $\alpha=0.1$ and $\alpha=0.05$. Snapshots of the
stationary state for these two values of $\alpha$ are shown in Figure
\ref{snaps}.
This figure reveals a striking similarity between synchronized regions
and a system with periodic boundary conditions and a narrow initial
distribution of $z$ values. Like the system with periodic boundary
conditions, the synchronized regions have different types of sites
that topple after 1, 2, 3 or 4 of their neighbors have toppled. The
probability distribution of force differences between neighbors has
peaks at multiples of $\alpha$ (figure not shown, but see for instance
\cite{gras94} for larger values of $\alpha$). Just as the pattern of
these different types of sites is generated by the initial system-wide
avalanche in the system with periodic boundary conditions,
synchronized regions are generated by large avalanches, which proceed
inward from the boundary. Sites that have participated in the same
avalanche have $z$ values within a limited range, and appear therefore
as a patch of a given grey shade in Fig.~\ref{snaps}. They remain in a
periodic state where the sites topple in the same order for a long
time. This can best be visualized by viewing the system after each
cycle, i.e., after a force increment of $1-4\alpha$. The inner part of
the system remains unchanged for many cycles. The mean time for which
a section of the system remains unchanged increases with increasing
system size, with decreasing $\alpha$, and with increasing distance
from the boundary.  If a section remains unchanged after a cycle, each
site has toppled once and has received a force increment of $4\alpha$
from its neighbors, just as in the system with periodic boundary
conditions, and often sites are lifted exactly to the threshold. The
periodic behavior of a synchronized region is terminated when an
avalanche enters it from outside and reshapes the pattern of $z$
values.

\begin{figure}
\includegraphics[width=8cm]{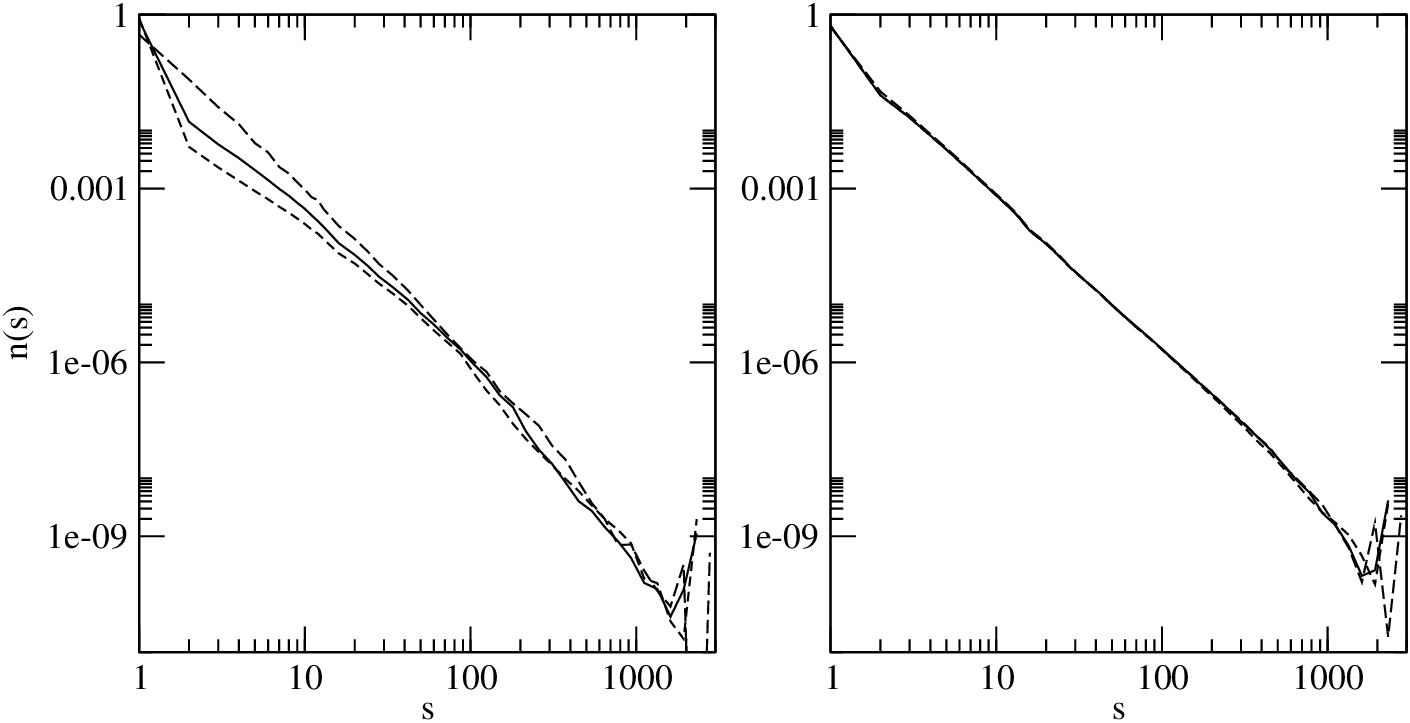}
\caption{Size distribution of avalanches for $L=100$ and
$\alpha=0.1$., averaged over approx.~200000 cycles. Left: Avalanches
triggered at least at a distance $L/4$ from the boundary. Right:
Avalanches triggered at most at a distance $L/8$ from the
boundary. Long dashed line: ``Float'' precision. Solid line:
``Double'' precision. Short dashed line: Curve obtained if sites that
are lifted exactly to the threshold do not participate in the
avalanche.
\label{ns1}}
\end{figure}

These similarities between synchronized regions and systems with periodic boundary conditions (and narrow initial width of $z$ values) suggest that the size distribution of avalanches is sensitive to the floating-point precision. This is illustrated in Fig.~\ref{ns1}.
In order to demonstrate the importance of the distance from the
boundary, avalanches triggered within the inner part of the system and
close to the boundary were monitored separately.  In addition to using
two different types of precision (``float'' and ``double''), a third
simulation was performed, where the $z$ value of sites that were
lifted exactly to the threshold by the force increment received from a
neighbor was slightly decreased (by $10^{-15}$) in order to prevent
such sites from participating in the avalanche. While for the lower
precision only 45 percent of all sites in the inner part of the system
topple in avalanches of size 1, this percentage increases to more than
80 percent with ``double'' precision. In the outer region, there is no
such sensitivity on the floating-point precision, due to the smaller
size of the regions and their shorter lifetime. 

\begin{figure}
\includegraphics[width=8cm]{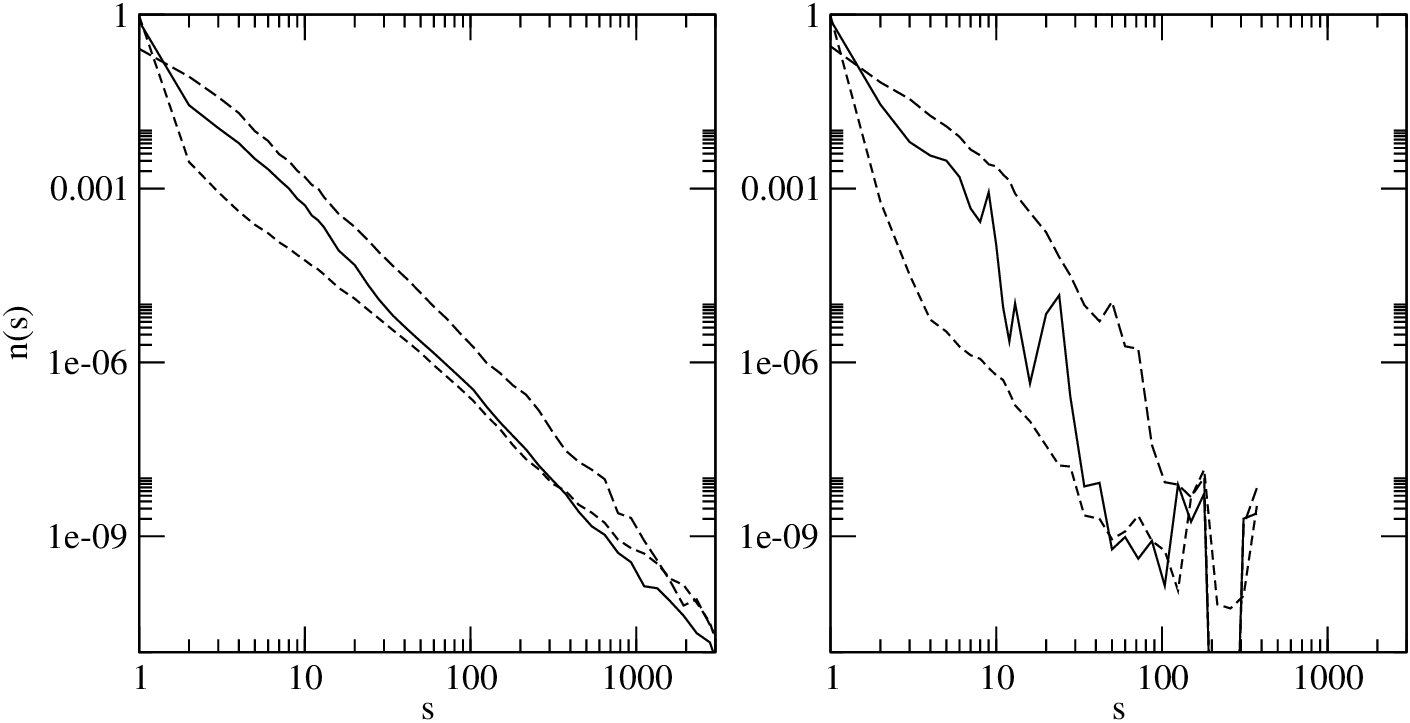}
\caption{Size distribution of avalanches for $L=200$ and
$\alpha=0.1$ (left) and $L=100$ and $\alpha=0.05$ (right), averaged over approx.~200000 cycles. Only avalanches
triggered at least at a distance $L/4$ from the boundary are considered.
Long dashed line: ``Float'' precision. Solid line:
``Double'' precision. Short dashed line: Curve obtained if sites that
are lifted exactly to the threshold do not participate in the
avalanche.
\label{ns2}}
\end{figure}

With increasing systems size and decreasing $\alpha$, the sensitivity to the floating-point precision increases, as shown in Fig.~\ref{ns2}.
With ``float'' precision, only 25 percent of the sites in the inner
region topple in avalanches of size 1 for $L=200$ and $\alpha=0.1$,
while this percentage is 75 for ``double'' precision and 94 when sites
that are lifted exactly to the threshold do not participate in the
avalanche. For even larger system size (and the same value
$\alpha=0.1$), on can expect the latter percentage to approach 100,
implying that virtually all topplings in the inner part of the system
occur in avalanches of size 1. A similar trend is observed for
decreasing $\alpha$. The curves in the right part of Fig.~\ref{ns2}
are not smooth due to the extremely long time scales for rearrangement
of the synchronized regions. Their cutoff occurs at smaller $s$ than
for $\alpha=0.1$, probably due to the small size of synchronized
regions.

Taking all these observations together, we can conclude that the
observed avalanche size distributions result from the superposition of
larger, synchronizing avalanches triggered near the boundaries and
smaller avalanches that occur within the synchronized regions and that
should turn into single-flip avalanches after a few cycles, if only
the floating-point precision was infinite. With increasing system size
and decreasing $\alpha$, most of the system stays for an increasingly
long time in the periodic state with single-flip avalanches. The
weight of larger avalanches would therefore decrease towards zero in
the thermodynamic limit of infinite system size, if only the
floating-point precision could be made infinite. The system is thus
not critical in the conventional sense that it has a scale-invariant
avalanche size distribution, with large and small avalanches being
generated by the same mechanism. Instead, the complex interplay of
synchronization, desynchronization, limited floating-point precision,
and a nontrivial size distribution of synchronized regions generates a
broad and power-law like avalanche size distribution for parameter
values typically used in simulations. Similar effects might be at the
bottom of many apparent power laws in nature.

\begin{acknowledgments}
This work was supported by the Deutsche 
Forschungsgemeinschaft (DFG) under Contract No Dr300-2/1.
\end{acknowledgments}

\bibliography{ofc}

\end{document}